# Introducing CitedReferencesExplorer (CRExplorer):

# A program for Reference Publication Year Spectroscopy

# with Cited References Standardization

Andreas Thor*, Werner Marx**, Loet Leydesdorff***, and Lutz Bornmann****

*University of Applied Sciences for Telecommunications Leipzig
Gustav-Freytag-Str. 43-45,
04277 Leipzig, Germany.
Email: thor@hft-leipzig.de

**Max Planck Institute for Solid State Research
Information Service
Heisenbergstrasse 1,
70506 Stuttgart, Germany.
Email: w.marx@fkf.mpg.de

***Amsterdam School of Communication Research (ASCoR),
University of Amsterdam,
P.O. Box 15793
1001 NG Amsterdam, The Netherlands.
Email: loet@leydesdorff.net

**** Corresponding author:
Division for Science and Innovation Studies
Administrative Headquarters of the Max Planck Society
Hofgartenstr. 8,
80539 Munich, Germany.
Email: bornmann@gv.mpg.de

**Abstract**

We introduce a new tool – the CitedReferencesExplorer (CRExplorer, www.crexplorer.net) – which can be used to disambiguate and analyze the cited references (CRs) of a publication set downloaded from the Web of Science (WoS). The tool is especially suitable to identify those publications which have been frequently cited by the researchers in a field and thereby to study for example the historical roots of a research field or topic. CRExplorer simplifies the identification of key publications by enabling the user to work with both a graph for identifying most frequently cited reference publication years (RPYs) and the list of references for the RPYs which have been most frequently cited. A further focus of the program is on the standardization of CRs. It is a serious problem in bibliometrics that there are several variants of the same CR in the WoS. In this study, CRExplorer is used to study the CRs of all papers published in the *Journal of Informetrics*. The analyses focus on the most important papers published between 1980 and 1990.

# 1  Introduction

The focus of bibliometric evaluation has been on the analysis of the citedness of publications (expressed as bare citation counts or normalized impact scores) (Vinkler, 2010). The times-cited data, however, is generated on the base of the cited references (CRs) which are given by the authors in their publications (Wouters, 1999). Although these CRs can be searched in Web of Science (WoS, Thomson Reuters), one usually does not work directly with CRs data, but instead with derived data, such as citation counts (i.e. the "times cited" assigned to the database records).

Comprehensive CR data have recently been successfully used to identify papers presenting interdisciplinary (Zhang, Rousseau, & Glänzel, 2015) and innovative research (Uzzi, Mukherjee, Stringer, & Jones, 2013). A specific kind of CR analysis is Reference Publication Year Spectroscopy (RPYS), which was introduced by Marx, Bornmann, Barth, and Leydesdorff (2014) formulating as follows: "RPYS is based on the analysis of the frequency with which references are cited in the publications of a specific research field in terms of the publication years of these CRs. The origins show up in the form of more or less pronounced peaks mostly caused by individual publications that are cited particularly frequently" (p. 751).

In the meantime, RPYS has been applied in a variety of contexts: Wray and Bornmann (2014) used RPYS to investigate the roots of the philosophy of science. They showed, among other things, that books play a more important role in philosophy of science than in the sciences. Furthermore, Einstein's (1905a; 1905b) papers created a considerable citation peak in this literature. In Marx et al. (2014) the origins of graphene and solar cells research were investigated and in a further publication the historical roots of the Higgs boson research were studied (Barth, Marx, Bornmann, & Mutz, 2014). Leydesdorff, Bornmann, Marx, and Milojevic (2014) studied the origins of "iMetrics" (information metrics, bibliometrics, and scientometrics) in scholarly literature. They showed that Lotka (1926) can be considered as the first common source of iMetrics, but that the intellectual program of iMetrics was shaped during the 1960s.

The results of Marx and Bornmann (2014) show that RPYS cannot only be used to investigate the historical roots of fields, but also to demask scientific legends: "Charles Darwin, the originator of evolutionary theory, was given credit for finches he did not see and for observations and insights about the finches he never made" (p. 839). The analysis validated the already well-known fact that the book of Lack (1947) is the origin of the term "Darwin finches". Comins and Hussey (2015a, 2015b) used RPYS to identify landmark research contributions to the global positioning system (GPS) and to investigate the impact of the Viterbi algorithm, first published by Andrew Viterbi in 1967.

Most RPYS papers published hitherto are based on software which can be downloaded at http://www.leydesdorff.net/software/rpys. In order to facilitate the use of RPYS, we developed a user interface which can be used both to analyze the CRs data in a set as well as to produce the results of the analyses in a graphical format for inclusion in a paper or presentation. The program is named CitedReferencesExplorer (CRExplorer) and can be found at www.crexplorer.net. The CRExplorer is aligned to and inspired by similar programs for analyzing and visualizing citation networks of publications (van Eck & Waltman, 2010; van Eck & Waltman, 2014).

Using all papers published in the *Journal of Informetrics* (JOI) as an example, we will present and discuss the CRExplorer in the following sections. Thus, the functionalities of the program will be explained by processing and analyzing the CRs from the papers published in JOI. A definitive focus will be on the functions of the program to standardize CRs. It is a serious problem in bibliometrics that there are several variants of the same CR in WoS. There are several issues in this context: CRs can refer to sources not belonging to the source items in WoS, journal name changes confuse authors in referencing, and typos on the side of the authors and the database providers (Thomson Reuters). The approach in the program to cluster and merge variants is pragmatic in trying to capture as much error as possible.

## 2      Methods

### 2.1    Example dataset

For the description of the CRExplorer, we generated a sample by downloading all papers published in *Journal of Informetrics* from WoS (date of reception: June 2015). This example dataset contains 545 papers which were downloaded in two packages (of 500 and 45 papers). The data contains 9,810 CRs. The dataset (two packages) can be downloaded from www.crexplorer.net (section guide & datasets).

### 2.2    The computer program CitedReferencesExplorer

CRExplorer is written in the Java programming language. Thus, the program runs on most hardware and operating system platforms. The program can freely be used.

## 3      Results

### 3.1    Working with a WoS dataset[1]

Using the menu item "File" – "Import WoS files", the CRExplorer opens one or several datasets from WoS (each download from WoS can contain up to 500 records). In WoS, the datasets are downloaded using the option "Save to Other File Formats". As "Record Content" select "Full Record and Cited References" and as "File Format" select "Other Reference Software". The records for the RPYS have to be searched in the WoS Core Collection in order to be able to save full records including the CRs. The CRExplorer imports maximally 100,000 CRs. This is the number which can be processed by most computers.

---

[1] The current version of the CRExplorer works with WoS data only. We plan for later versions that also Scopus data can be processed.

However, the user can change this number using "File" – "Settings" – Import". Setting the number to zero means that there is no import limit, but the processing is limited by the available memory on the computer.

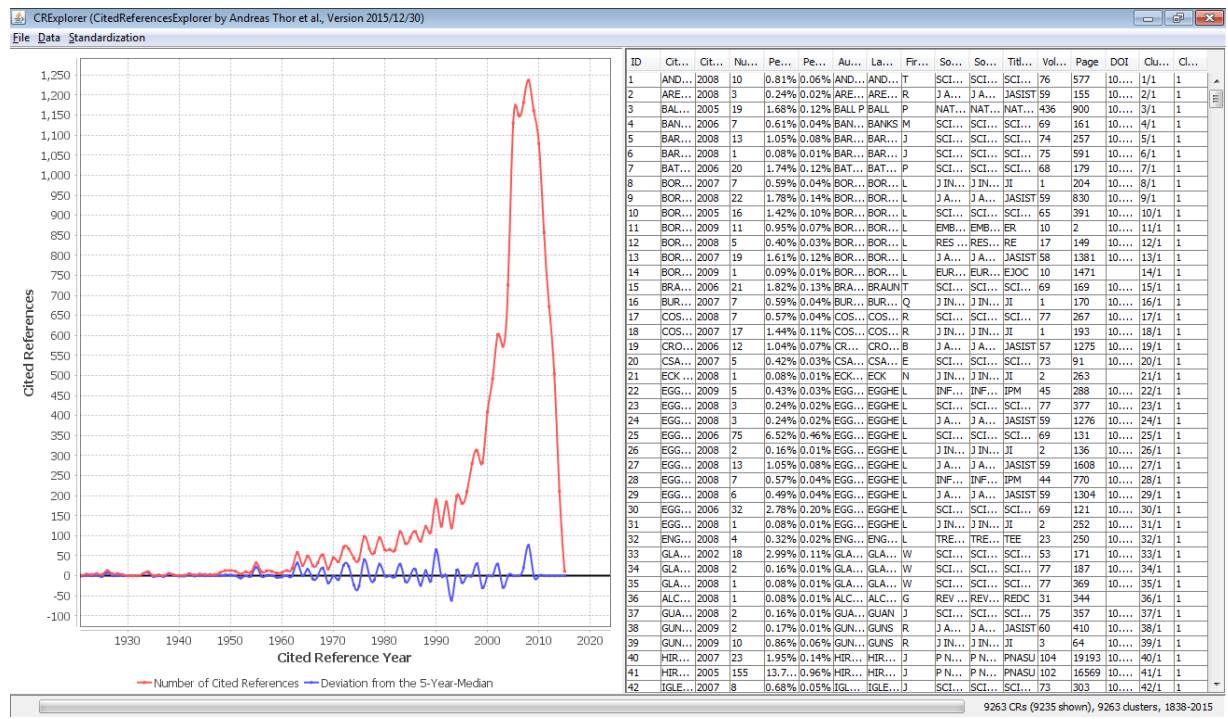

Figure 1. Screenshot of the CitedReferencesExplorer after loading the example data.

In some cases, the user wants to import large datasets (e.g. all publications on a specific topic or from a broader research field) in order to investigate the historical roots. When CRExplorer is not able to process the complete bibliographic records (because of computer memory restriction), the user has the possibility to restrict the imported range of cited references years (CRYs) using "File" – "Settings" – "Import" – "Minimum/ Maximum Publication Year of Cited References". Thus, it is possible to investigate the historical roots of topics, fields, institutions etc. based on large datasets by only considering the CRYs which are the target of investigation. The default in "Minimum/ Maximum Publication Year of Cited References" is zero which means that the CRY is not relevant for the import. The zero can also be used to define the import range (e.g. Minimum=1900 and Maximum=0 mean that all CRs with CRY>=1900 are imported).

After uploading the example dataset of 545 papers (saved in two packages), a message box appears providing the number of papers, the number of CRs uploaded, and some more information. This information is also available at the menu item "Data" – "Info"; the status bar informs constantly about the number of CRs, the number of clusters (see section 3.3), and the range of RPYs. The screen of the CRExplorer has two parts (see a screenshot in Figure 1): On the left side, the number of CRs per CRY is visualized (red line, "Number of Cited References"). The blue line in the graph ("Deviation from the five-year Median") shows the deviation of the number of CRs in each year from the median for the number of CRs in the two previous, the current, and the two following years ($Y-2; Y-1; Y; Y+1; Y+2$). This deviation from the five-year median provides a curve smoother than the one in terms of absolute numbers. If there is no CR in a specific RPY, the RPY is set to zero for the calculation of the deviation from the median. Using the smoothed curve, peaks in the data can be identified more easily than with the number of CRs curve, since each year is compared with its adjacent years. The user of the CRExplorer can change the number of two to any other number and can thus work with medians calculated based on a different time window instead of a five-year time window.

Choosing "File" – "Settings" – "Chart" the user can select the lines which should be displayed: "Number of Cited References", "Deviation from the Median" or both. The width of the chart lines and the median range for the smoothed curve can also be changed (at "File" – "Settings" – "Chart"). Further options for changing (e.g. the axis labels or legend) and saving the figure can be found by right-clicking on the figure.

On the right side, a table lists all CRs included in the RPYS: The table shows the CR as found in the WoS data (column "Cited Reference"), the year when the CR was published (column "Cited Reference Year"), and the number of times a specific reference has been cited (column "Number of Cited References"). Every CR in the table receives a unique identification number (column "ID"). Furthermore, two percentages are given: "Percent in year" is the proportion of the number of times a specific reference has been cited among the number of all CRs in the same reference publication year. "Percent over all years" is the proportion using all CRs over all reference publication years.

Further columns contain splits of the data in the CR: "Author", "Last name", "First name initial", and "Source" (showing mostly journal title, volume, issue, and first page). "Source title" contains mostly the journal title or the abbreviated book title; "Title short" contains the first letters of the words in "Source title" if there is more than one word. If there is only a single word, this word appears in "Title short", too. "Volume", "Page", and "DOI" show the corresponding data (if available in the WoS files) if the CR is a journal paper. Two further columns "ClusterID" and "Cluster Size" will be explained in section 3.3.

The data in the columns of the table can be sorted in ascending or descending order. For example, sorting the example dataset by "Number of Cited References" shows that the most-cited paper in the set is the paper of Hirsch (2005) with 171 citations; this paper represents nearly 1% of all the references provided in JOI. It is also possible to sort by multiple columns: For example, if one wants to sort by "Percent in Year" and then by "Number of Cited References", click firstly on "Number of Cited References" and then on "Percent in Year". Sorting initializes the background color pattern of the table rows: Rows with the same value in the sorted columns have the same background color. The pattern supports the inspecting of the CRs data by highlighting the rows with the same value.

Choosing "File" – "Settings" – "Table" the user can select those columns which should be displayed in the table. This function enables to restrict the columns to those which are needed for a specific analysis.

Our intention is that the user of the CRExplorer is enabled to work with the graph on the left side of the screen to analyze the data. To this purpose, a number of facilities have been developed (which we intend to extend on the basis of feedback by the users):

(1) Sweeping with the mouse over a data point on one of the curves in the graph calls a small window showing the corresponding CRY ("Year") and the sum of the CRs in this year ("N_CR"). Thus, the years of peaks in the curves and their impact can be identified easily.

(2) Clicking on a data point, the CRs data in the table (on the right side of the screen) is sorted by "Cited Reference Year" and "Percent in year" (in descending order). Furthermore, the first CR with the highest percentage in the particular year is marked by a blue line. Since the data is sorted by the "Percent in Year", one can inspect the most important CRs which are accountable for a peak.

In the example dataset, for example, a peak is visible at the year 1955. The click on the deviation from median curve shows that Eugene Garfield's (1955) well known article in *Science* introducing citation indexes is the most CR in this year (25 citations corresponding to nearly 73.5% of all citations in this year). Two earlier peaks are also visible for the CRYs 1926 and 1934 which can be traced back to Lotka (1926) and Bradford (1934) (Leydesdorff et al., 2014). Whereas Bradford (1934) is responsible for 80% of the references in this year, 100% of the references to the year 1926 can be traced back to Lotka (1926).

The paper of Lotka (1926) provides an example of the general problem with CR data: several variants exist for one and the same publication (e.g., "lotka a.j., 1926, j washington acad sc, v16, p317" and "lotka aj, 1926, j washington acad sc, p292"). We return to this problem (the aggregation of variants) in section 3.3.

(3) The example dataset includes CRs from 1781 until 2015. Since this is a very long time period, it can be difficult to inspect the CRYs and to search for peaks – especially in early years. Using the mouse, one can mark an area on the graph or, alternatively, a time period can be selected by using the menu item "Data" – "Filter by Cited References Year". Thereafter, the graph is restricted to the marked period (see also the following chapter 3.2).

Figure 2 shows a selection from the example dataset (using only CRYs from 1900 to 1960). This selection highlights the importance of the three above mentioned publications (Bradford, 1934; Garfield, 1955; Lotka, 1926). Further peaks are visible, for example, for the year 1914. In this year, Hazen (1914) published a formula which is important for the calculation of percentiles of citations (Bornmann, Leydesdorff, & Mutz, 2013). One can re-size the vertical (range) axis by right-clicking on the figure and following the options provided.

The user can recall the initial graph (and thus dissolve the selection) by right-clicking on the graph and choosing "auto-range" from the menu.

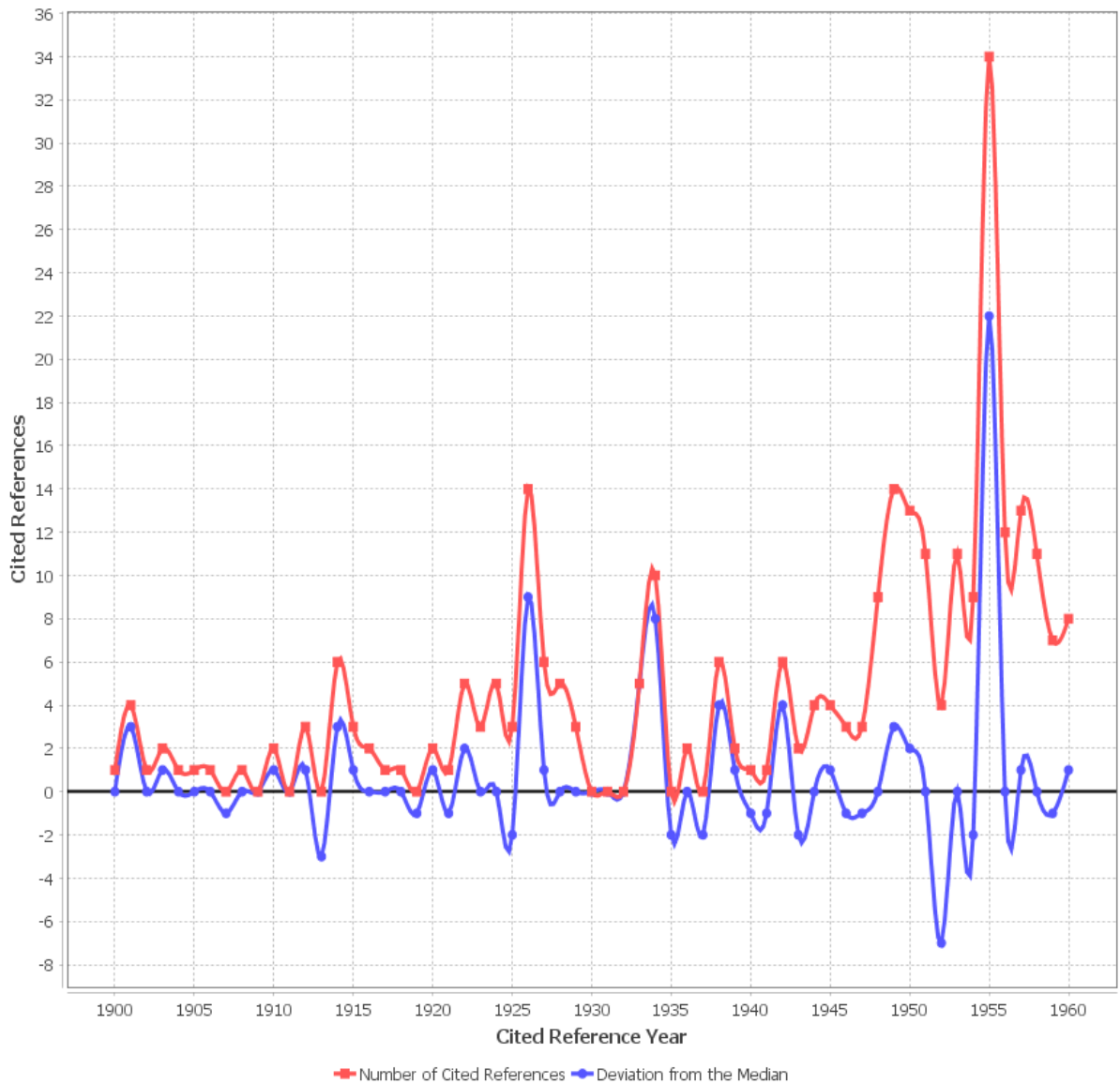

Figure 2. Reference Publication Year Spectroscopy for the *Journal of Informetrics* (restricted to cited reference years from 1900 to 1960)

### 3.2 Removing data from a dataset

The analyst may often wish to restrict RPYS to a certain time period (such as 1900-1960 in the example above). Very early and most recent years are frequently not very helpful for the identification of the most cited publications in the history (Comins & Leydesdorff, 2016). Thus, most RPYSs published hitherto removed data from the analyses in order to present a final result (e.g., Figure 2 above).

Although the CRExplorer allows for the selection of periods in the graph, this selection does not lead to changes in the column of table "Percent over all years" on the right side of the screen. However, there are four ways of removing data from a dataset which are explained in the following:

(1) Lines in the table on the right side of the screen can be marked and deleted using the menu item "Data" – "Remove selected Cited References".

(2) Selecting "Data" – "Remove by Cited Reference Year" the user can remove the data for specific CRYs. Figure 3 shows the result of two removals for the time periods from 1781 until 1959 and from 1981 until 2015. The first and last CRYs in a dataset can be identified by sorting the data in the table by "Cited Reference Year" or by selecting "Data" – "Info". Thus, the figure shows the results for the time period from 1960 until 1980.

The most pronounced peak is visible for 1976 with Pinski and Narin (1976) and de Solla Price (1976) as the most important publications. These two papers are responsible for approximately 40% of the references for this year. Pinski and Narin (1976) is often cited as the first study with a recursive algorithm for measuring the influence of citations. A modification of this algorithm is nowadays used at the internet as PageRank (Page, Brin, Motwani, & Winograd, 1998), and for journals in the cases of SJR (Gonzalez-Pereira, Guerrero-Bote, & Moya-Anegon, 2010; Guerrero-Bote & de Moya-Anegon, 2012) and Eigenfactor score (Bergstrom, 2007).

A second peak is visible for the year 1963 which is mostly produced by referencing Derek de Solla Price's (1963) book "Little science, big science". This book can be considered as the first in a series of studies of this author that led to the emergence of the scientometric research program (e.g. de Solla Price, 1976). Inspection of the results for 1963 shows again the problem of name variants of one and the same publication. de Solla Price's (1963) book appears in the list of CRs with four variants: "price

d.j.d.s., 1963, little sci big sci", "de solla price derek j, 1963, little sci big sci",

"[anonymous], 1963, little sci big sci", and "price, 1963, little sci big sci".

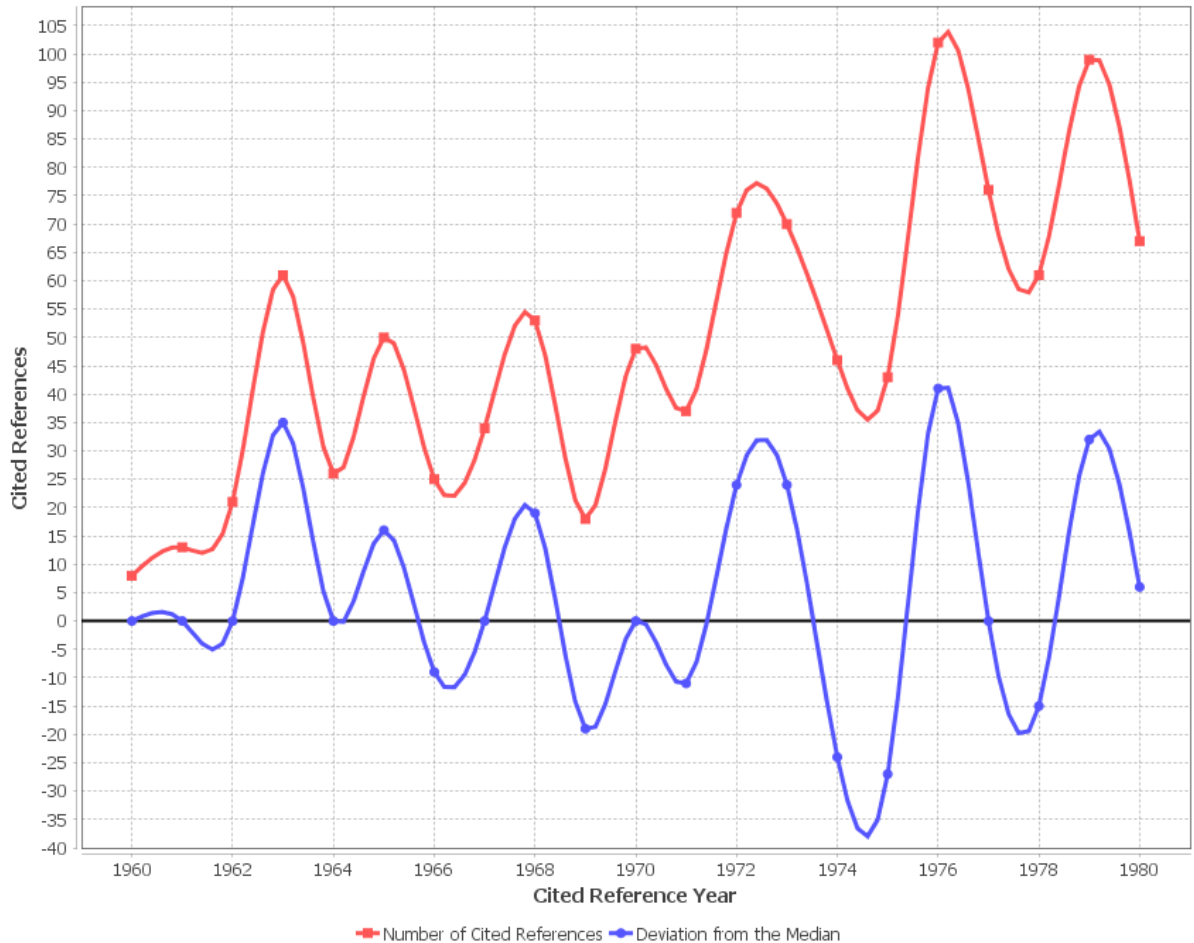

Figure 3. Reference Publication Year Spectroscopy for the *Journal of Informetrics* (restricted to cited reference years from 1960 to 1980).

(3)    A third option for removing data is by removing CRs below a minimum number of CRs. This can be done by using the menu item "Data" – "Remove by Number of cited references". Using this option, we removed from the example dataset all CRs with less than 10 citations (the threshold of 10 is arbitrary and others can also be used); the corresponding graph is shown in Figure 4. This kind of restriction is helpful in identifying publications from early CRYs with a substantial impact (and to suppress the noise of less cited publications). Furthermore, in CRYs with many sparsely cited publications the publications with substantial impact can then be easier identified. The most pronounced peak in Figure 4 is for the CRY 2006; with nearly 20% of the references, this peak is dominantly shaped by Egghe's (2006) studies introducing the g-index.

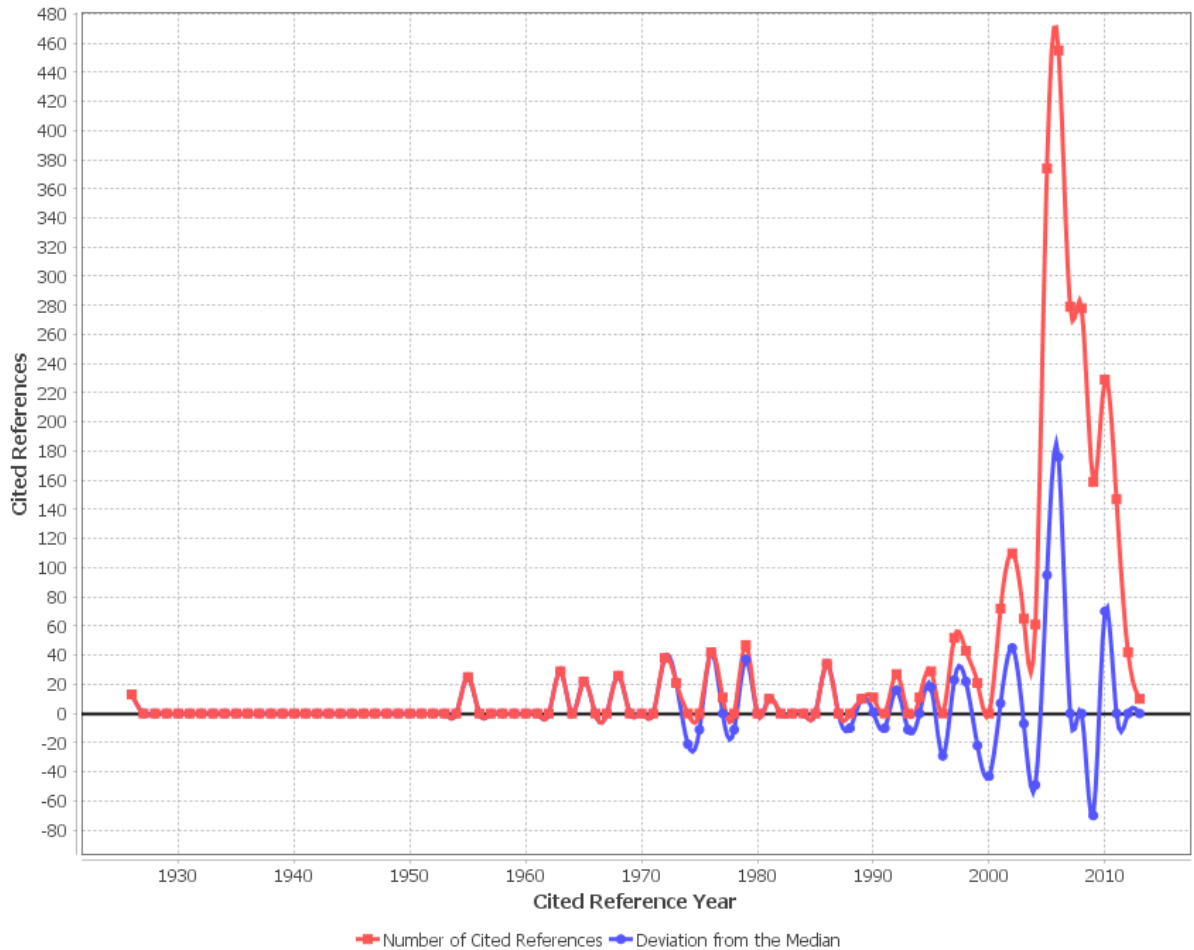

Figure 4. Reference Publication Year Spectroscopy for the *Journal of Informetrics* (restricted to cited references with at least 10 citations).

(4) The fourth option for removing data is similar to the third: CRs can be removed by using thresholds for the field "Percentage in Year". Thus, it is possible to remove lowly cited references whereby "lowly" is defined in terms of the citation distribution in the CRYs.

The graph on the left side of the screen can be printed, copied and saved in various formats (by right-clicking on the graph). Furthermore, the properties of the graph can be considerably changed here: the title can be deleted or re-worded; the labels of the x- and y-axes can be changed, etc. Thus, the user is able to adapt the graph to his/ her needs for showing the graph (in a paper or presentation). The graph data can also be exported for further processing in programs such as Excel (save the graph as csv file then).

The table on the right side of the screen can also be saved as a csv-file ("File" – "Save as CSV File"). Thus, the user can incorporate the table (or a selection of the table) in a paper with the results of RPYS. Furthermore, the user can save a specific work result: The changed dataset (initially from WoS) can be saved and opened later using "File" – "Open CSV File".

### 3.3 Eliminating variants of cited references in a WoS dataset

In the previous sections, we already mentioned the problem of variants of the same CR potentially disturbing the analysis. At the menu items "Standardization" – "Cluster equivalent Cited References" / "Merge Cited References of the Same Cluster" the user has the possibility to detect variants of the same CR, cluster them, and merge their occurrences (number of CRs). Note that the automatic clustering of variants is restricted to variants within the same publication year and not across publication years. Thus, a reference to a first edition of a book and a reference to a later edition are not clustered by this routine. However, the user can make further adjustments to the author names, the journal or book title and other bibliographic information of CRs (across publication years). The clustering uses the table with the list of CRs on the right side of the screen as input file. Since the clustering (and merging) is a complex process which needs a lot of computer resources, the user should cluster only those data which is of interest. For example, if the user is interested in results for the CRYs 1970 and earlier, s/he is advised to restrict the dataset to the years subsequent to the start of the clustering and merging process.

In the first step of eliminating variants of CRs, the variants of the same cited publication are identified. When the user selects "Standardization" – "Cluster equivalent Cited References", two attributes are used for a first similarity computation: "Last name" of the author and "Source title" from the table on the right side of the screen. Based on this data, the CRExplorer determines the pair-wise similarity of variants of CRs. The program computes the Levenshtein similarity (as provided by the SimMetrics library https://github.com/Simmetrics/simmetrics) of both attributes (see also Wasi & Flaaen, 2015). The Levenshtein *similarity* of two strings s1 and s2 is defined as sim(s1,s2) = 1-LD(s1,s2)/max(|s1|,|s2|). Here |s| denotes the length of a string s and LD (s1, s2) is the Levenshtein *distance* which is defined as follows: The Levenshtein distance between two strings s1 and s2 is the minimal number of single-character edit operations (i.e., insertion, deletion, or substitution) required to transform string s1 into s2. The Levenshtein distance is 0 for equal strings (no edit operations necessary) and equals max(|s1|,|s2|) for totally different strings (substitute the first min(|s1|,|s2|) characters and insert / delete the remaining characters). Therefore, for any two strings the Levenstein *similarity* is between 0 to 1 where 0 corresponds to "totally different" and 1 to "identical".

For two CRs $o_1$ and $o_2$, CRExplorer computes the Levenshtein similarity of the first authors' last names as well as the similarity of the source titles. The two CRs $o_1$ and $o_2$ are considered as "matching" if the weighted average (ratio 2:1) of the two similarity values is equal to or greater than the threshold of 0.75 (this can be changed by the user, see below). The combination of multiple similarity values that are based on different attributes typically achieves a better match quality compared to a single similarity of the entire CR strings. First, it restricts the similarity computation to relevant (and available) attribute values. Second, the combination allows for an appropriate weighting of attributes independent from their actual string length.

The CRExplorer performs a clustering based on the matching results, i.e., the list of the matching CR pairs. Two CRs $o_1$ and $o_2$ are assigned the same cluster, if the pair $(o_1,o_2)$ appears in the matching result or if there is a list of other CRs $t_1, ..., t_n$ so that $(o_1,t_1)$, $(t_1,t_2)$, $(t_2,t_3)$, ..., $(t_{n-1},t_n)$, and $(t_n, o_2)$ are all among the matching pairs. Each cluster is uniquely identified by its ClusterID, i.e., all CRs of a cluster are marked with a corresponding ClusterID. Thus, the results of the similarity computation can be inspected using the column "ClusterID" in the table. The number of CRs in each cluster is provided by the column labeled "Cluster Size".

We used the example dataset in order to cluster variants of the same CR. For example, Table 1 shows variants of one CR (Hirsch, 2005) in the dataset. These variants received the same ClusterID by the program (here: 41/41) when the program has finished the similarity computation. The program works with two ID numbers divided by a slash to facilitate a manual post-correction if the clustering led to erroneous results. The process of manual post-correction will be described later on.

In the second step of eliminating variants of CRs (subsequent to the clustering of variants), the menu item "Standardization" – "Merge Cited References of the same cluster" starts the process where "ClusterID" is used for the aggregation, i.e., the values of the corresponding lines in the table on the right sight of the screen are summarized per cluster. Thus, the numbers of citations for the seven variants in Table 1 are summarized to n=177. From the variants (CRs) forming a cluster the one variant is chosen as representative which has the highest number of citations in the cluster ("Number of Cited References"). In the (unlikely) case of multiple variants in a cluster with a maximum number of citations, the choice is random. At the example in Table 1, the representative "Hirsch JE, 2005, P NATL ACAD SCI USA, V102, P16569, DOI 10.1073/pnas.0507655102" is chosen, because it was cited 171 times; all other variants are single occurrences.

Table 1. Variants of the same cited reference (Hirsch, 2005) in the example dataset which are clustered

| Cited reference | ClusterID | Number of citations |
|---|---|---|
| HIRSCH J, 2005, P NATL ACAD SCI USA, P16569 | 41/41 | 1 |
| Hirsch J., 2005, P NATL ACAD SCI USA, V102, P165 | 41/41 | 1 |
| Hirsch J. E, 2005, P NATL ACAD SCI USA, V102, P16569 | 41/41 | 1 |
| Hirsch J. E., 2005, P NATL ACAD SCI, V102, P16569 | 41/41 | 1 |
| Hirsch J. E., 2005, P NATL ACAD SCI USA, V102, P16569 | 41/41 | 1 |
| Hirsch JE, 2005, P NATL ACAD SCI USA, V102, P16569, DOI 10.1073/pnas.0507655102 | 41/41 | 171 |
| Hirsch JE, 2005, P NATL ACAD SCI USA, V102, P16572, DOI DOI 10.1073/PNAS.0507655102 | 41/41 | 1 |
| Total | | 177 |

The clustering procedure ("Standardization" – "Cluster equivalent Cited References") can be helpful in aggregating variants of the same CR. However, this procedure is prone to error itself. For example, if there are several CRs from the same authors and published in the same journal in one year, these CRs are clustered, although they refer to different publications. This error affects journal papers in most of the cases. Table 2 shows several CRs which received by the procedure the same ClusterID (using the example dataset). However, the CRs refer to 5 different publications and should not be clustered (merged).

Table 2. Different cited papers in the example dataset which are clustered erroneously. The papers were published by P. Jacso in *Online Information Review* in the same year (2008).

| Cited reference | ClusterID | Number of citations |
|---|---|---|
| Jacso P, 2008, ONLINE INFORM REV, V32, P266, DOI 10.1108/14684520810879872 | 44/44 | 4 |
| Jacso P, 2008, ONLINE INFORM REV, V32, P437, DOI 10.1108/14684520810889718 | 44/44 | 4 |
| Jacso P, 2008, ONLINE INFORM REV, V32, P102, DOI 10.1108/14684520810866010 | 44/44 | 3 |
| Jacso P, 2008, ONLINE INFORM REV, V32, P524, DOI 10.1108/14684520810897403 | 44/44 | 2 |
| Jacso P, 2008, ONLINE INFORM REV, V32, P673, DOI 10.1108/14684520810914043 | 44/44 | 2 |

Thus, we strongly recommend that a user of the procedure controls the results from the clustering procedure ("Standardization" – "Cluster equivalent Cited References") and corrects wrong matches. For the manual correction we have implemented some features in the program which support the user in post-processing the clustered results. Corresponding control buttons appear above the table on the right side of the screen, when the user starts the clustering process (by choosing "Standardization" – "Cluster equivalent Cited References"). We will illustrate the features based on clustered CRs in the example dataset.

The seven features can be used separately or in combination to change the cluster results before the merging of the CR variants is started. The effects of the features can be inspected by the values in "ClusterID" which has two components: The first value in the column (before the slash) shows the cluster numbers which result from the initial clustering process which was done automatically. The second value (after the slash) marks sub-clusters which change after using the features. Thus, the user should inspect the second value of the ClusterID to assess the results of the chosen post processing.

The seven features implemented in the program are the following:

1) **Levenshtein**: For matching similar CRs, Levenshtein is initially used as similarity function with a threshold of 0.75. However, the user can change this afterwards by using the slide control which accepts values between 0.5 (shown as 50) and 1 (shown as 100). If the slider is moved in the direction of 50, less similar CRs are matched; by moving the slider in the direction of 100, the matching process becomes increasingly restricted.

    For example, if the Levenshtein threshold of 0.5 is set by the user, more CRs appear with the ClusterID 44/44. Whereas the threshold of 0.75 leads to five possible variants of the same CR (see Table 2), the threshold of 0.5 results in eight variants (see Table 3). In this case, the reduction of the threshold does not make sense, because more CRs are erroneously clustered.

Table 3. Different cited references in the example dataset which are clustered (erroneously) using the Levenshtein threshold of 0.5. The papers were published by P. Jacso and M. O. Jackson in 2008.

| Cited reference | ClusterID | Number of citations |
| --- | --- | --- |
| Jacso P, 2008, ONLINE INFORM REV, V32, P266, DOI 10.1108/14684520810879872 | 44/44 | 4 |
| Jacso P, 2008, ONLINE INFORM REV, V32, P437, DOI 10.1108/14684520810889718 | 44/44 | 4 |
| Jacso P, 2008, LIBR TRENDS, V56, P784 | 44/44 | 5 |
| Jacso P., 2008, GOOGLE SCHOLAR SCI | 44/44 | 2 |
| Jacso P, 2008, ONLINE INFORM REV, V32, P102, DOI 10.1108/14684520810866010 | 44/44 | 3 |
| Jackson MO, 2008, SOCIAL AND ECONOMIC NETWORKS, P1 | 44/44 | 1 |
| Jacso P, 2008, ONLINE INFORM REV, V32, P524, DOI 10.1108/14684520810897403 | 44/44 | 2 |
| Jacso P, 2008, ONLINE INFORM REV, V32, P673, DOI 10.1108/14684520810914043 | 44/44 | 2 |

2) **Volume**, **Page**, and **DOI**: The user can select volume, page, and DOI in order to differentiate the clusters further. These selections affect the whole dataset and not only CRs which are marked by the user. Note that the Levenshtein approach is not applied to volume, page, and DOI; a perfect match is required for these attributes. For example, after selecting "Page" in the control panel all CRs in Table 3 are separated as different sub-clusters (see Table 4). In the cases under study, the consideration of the page number leads to the needed elimination of the cluster solution suggested by the program.

Table 4. New ClusterIDs for the cited references in Table 3. The new ClusterIDs appeared after selection of "Page".

| Cited reference | ClusterID | Number of citations |
| --- | --- | --- |
| Jacso P, 2008, ONLINE INFORM REV, V32, P266, DOI 10.1108/14684520810879872 | 44/44 | 4 |
| Jacso P, 2008, ONLINE INFORM REV, V32, P437, DOI 10.1108/14684520810889718 | 44/45 | 4 |

| Cited reference | | |
|---|---|---|
| Jacso P, 2008, LIBR TRENDS, V56, P784 | 44/46 | 5 |
| Jacso P., 2008, GOOGLE SCHOLAR SCI | 44/1274 | 2 |
| Jacso P, 2008, ONLINE INFORM REV, V32, P102, DOI 10.1108/14684520810866010 | 44/2001 | 3 |
| Jackson MO, 2008, SOCIAL AND ECONOMIC NETWORKS, P1 | 44/4907 | 1 |
| Jacso P, 2008, ONLINE INFORM REV, V32, P524, DOI 10.1108/14684520810897403 | 44/5071 | 2 |
| Jacso P, 2008, ONLINE INFORM REV, V32, P673, DOI 10.1108/14684520810914043 | 44/5072 | 2 |

3) **Manual generation of sub-clusters**: The tool offers three different ways for the manual changing of sub-clusters. They are named as "Same", "Different", and "Extract". Most of the problems with the automated clustering procedure occur with false positives: The algorithm matches CRs, although they should be separated. For the manual separation of clusters, the user can apply "Different" or "Extract". "**Different**" assigns different sub-cluster IDs to those CRs which are marked by the user manually (using the mouse click). "**Extract**" puts the marked CRs in a separated sub-cluster. "**Same**" gives marked CRs the same sub-cluster-ID. Manual changes based on "Same", "Different", and "Extract" can be rolled back using the "**Undo**" button.

We will illustrate the manual generation of sub-clusters in the following. Table 5 shows a cluster of CRs which was automatically generated by the CRExplorer.

Table 5. ClusterIDs of CRs before and after the manual correction

| Cited reference | ClusterID | Number of citations |
|---|---|---|
| Before manual correction | | |
|     Schreiber M, 2012, J AM SOC INF SCI TEC, V63, P2062, DOI 10.1002/asi.22703 | 2487/3730 | 2 |
|     Schreiber M, 2012, J INFORMETR, V6, P347, DOI 10.1016/j.joi.2012.02.001 | 2487/3730 | 2 |
| After manual correction | | |
|     Schreiber M, 2012, J AM SOC INF SCI TEC, V63, P2062, DOI 10.1002/asi.22703 | 2487/3730 | 2 |

| Cited reference | | |
|---|---|---|
| Schreiber M, 2012, J INFORMETR, V6, P347, DOI 10.1016/j.joi.2012.02.001 | 2487/5884 | 2 |

Table 5 shows ClusterIDs of CRs before and after the manual correction. The program clustered two CRs (Schreiber, 2012; Schreiber, Malesios, & Psarakis, 2012); but these should be separated. By marking the second CR (Schreiber M, 2012, J INFORMETR, V6, P347, DOI 10.1016/j.joi.2012.02.001) and clicking on "Extract", this CR is no longer part of the cluster formed by the other CRs.

Table 6. ClusterIDs of cited references before and after manual corrections

| Cited reference | ClusterID | Number of citations |
|---|---|---|
| Before manual correction | | |
| Leydesdorff L, 2008, J AM SOC INF SCI TEC, V59, P1582, DOI 10.1002/asi.20814 | 54/54 | 1 |
| Leydesdorff L, 2008, J AM SOC INF SCI TEC, V59, P1810, DOI 10.1002/asi.20891 | 54/54 | 6 |
| Leydesdorff L, 2008, J AM SOC INF SCI TEC, V59, P278, DOI 10.1002/asi.20743 | 54/54 | 10 |
| Leydesdorff L, 2008, J AM SOC INF SCI TEC, V59, P77, DOI 10.1002/asi.20732 | 54/54 | 2 |
| Leydesdorff L, 2008, J INFORMETR, V2, P317, DOI 10.1016/j.joi.2008.07.003 | 54/54 | 6 |
| Leydesdorff L., 2008, J AM SOC INFORM SCI, V591, P1810 | 54/54 | 1 |
| After using "Different" for the cited references | | |
| Leydesdorff L, 2008, J AM SOC INF SCI TEC, V59, P1582, DOI 10.1002/asi.20814 | 54/7857 | 1 |
| Leydesdorff L, 2008, J AM SOC INF SCI TEC, V59, P1810, DOI 10.1002/asi.20891 | 54/863 | 6 |
| Leydesdorff L, 2008, J AM SOC INF SCI TEC, V59, P278, DOI 10.1002/asi.20743 | 54/54 | 10 |
| Leydesdorff L, 2008, J AM SOC INF SCI TEC, V59, P77, DOI 10.1002/asi.20732 | 54/2588 | 2 |
| Leydesdorff L, 2008, J INFORMETR, V2, P317, DOI 10.1016/j.joi.2008.07.003 | 54/943 | 6 |
| Leydesdorff L., 2008, J AM SOC INFORM SCI, V591, P1810 | 54/1950 | 1 |
| After using "Same" for the marked cited references | | |
| Leydesdorff L, 2008, J AM SOC INF SCI TEC, V59, P1582, DOI 10.1002/asi.20814 | 54/7857 | 1 |
| Leydesdorff L, 2008, J AM SOC INF SCI TEC, V59, | 54/863 | 6 |
| P1810, DOI 10.1002/asi.20891 | | |
| Leydesdorff L, 2008, J AM SOC INF SCI TEC, V59, P278, DOI 10.1002/asi.20743 | 54/54 | 10 |
| Leydesdorff L, 2008, J AM SOC INF SCI TEC, V59, P77, DOI 10.1002/asi.20732 | 54/2588 | 2 |
| Leydesdorff L, 2008, J INFORMETR, V2, P317, DOI 10.1016/j.joi.2008.07.003 | 54/943 | 6 |
| Leydesdorff L., 2008, J AM SOC INFORM SCI, V591, P1810 | 54/863 | 1 |

In Table 6 another case of manual correction is shown. The program has clustered six cited papers of L. Leydesdorff, although only two should be clustered (see the grey marked CRs in the table) and the others not. The erroneous clustering can be corrected in two steps: In the first step all CRs in the cluster are marked and each reference is assigned a different ClusterID by clicking on "Different". In the second step, the CRs which are grey marked in Table 6 receive the same ClusterID (54/863), after marking by the user and the button "Same" is clicked. After this step, the two CRs in the table are merged.

### 3.4 Most frequently cited references from authors in JOI between 1980 and 1990

Using the automatic and manual clustering and merging functionalities of the CRExplorer, we produced a CR set based on the example data set where all variants of the same CRs are merged. The initial set of CRs from the JOI papers consists of n=9,810 CRs in total. We restricted the clustering and merging process to n=785 CRs from the CRYs 1980 to 1990. Using the automatically and manually clustering and merging functionalities of the program this set is reduced to n=750 CRs.

In a first step after finishing the clustering process by the program, we selected "Volume" and "Page" in order to separate the CRs further on and thus aggregated 22 CRs. In a second step, the CRs were arranged alphabetically and 18 CRs were manually changed by using "Same". The alphabetical order of the CRs was needed to reveal, for example, misspelled reference variants (mostly inverted volume and page numbers). All dubious variants of one and the same author and publication year were checked using the WoS database.

Table 7. Most frequently cited publications from authors published in JOI between 1980 and 1990

| Cited reference | Number of citations | Proportion of citations within a year |
|---|---|---|
| 1980 | | |
| (1) FREEMAN LC, 1980, QUAL QUANT, V14, P585 | 4 | 5.97 |
| (2) LINDSEY D, 1980, SOC STUD SCI, V10, P145, DOI 10.1177/030631278001000202 | 4 | 5.97 |
| 1981 | | |
| (3) WHITE HD, 1981, J AM SOC INFORM SCI, V32, P163, DOI 10.1002/asi.4630320302 | 10 | 14.93 |
| (4) HODGE SE, 1981, SCIENCE, V213, P950 | 7 | 10.45 |
| 1982 | | |
| (5) DOSI G, 1982, RES POLICY, V11, P147, DOI 10.1016/0048-7333(82)90016-6 | 3 | 4.29 |
| 1983 | | |
| (6) SCHUBERT A, 1983, SCIENTOMETRICS, V5, P59, DOI 10.1007/BF02097178 | 9 | 7.83 |
| (7) CALLON M, 1983, SOC SCI INFORM, V22, P191, DOI 10.1177/053901883022002003 | 9 | 7.83 |
| (8) MARTIN BR, 1983, RES POLICY, V12, P61, DOI 10.1016/0048-7333(83)90005-7   1983 | 9 | 7.83 |
| 1984 | | |
| (9) HAUSMAN J, 1984, ECONOMETRICA, V52, P909, DOI 10.2307/1911191 | 7 | 7.87 |
| (10) Cronin B., 1984, CITATION PROCESS ROL | 7 | 7.87 |
| 1985 | | |
| (11) MOED HF, 1985, RES POLICY, V14, P131, DOI 10.1016/0048-7333(85)90012-5 | 9 | 9.18 |
| (12) SMALL H, 1985, SCIENTOMETRICS, V7, P391, DOI 10.1007/BF02017157 | 8 | 8.16 |
| 1986 | | |
| (13) SCHUBERT A, 1986, SCIENTOMETRICS, V9, P281, DOI 10.1007/BF02017249 | 22 | 18.64 |
| (14) VINKLER P, 1986, SCIENTOMETRICS, V10, P157, DOI 10.1007/BF02026039 | 12 | 10.17 |
| 1987 | | |
| (15) SCHUBERT A, 1987, SCIENTOMETRICS, V12, P267, DOI 10.1007/BF02016664 | 8 | 9.09 |
| (16) BONACICH P, 1987, AM J SOCIOL, V92, P1170, | 7 | 7.95 |
| DOI 10.1086/228631 | | |
| 1988 | | |
| (17) Cohen J, 1988, STAT POWER ANAL BEHA | 9 | 6.77 |
| (18) GLANZEL W, 1988, J INFORM SCI, V14, P123, DOI 10.1177/016555158801400208 | 6 | 4.51 |
| 1989 | | |
| (19) KAMADA T, 1989, INFORM PROCESS LETT, V31, P7, DOI 10.1016/0020-0190(89)90102-6 | 11 | 9.24 |
| (20) MACROBERTS MH, 1989, J AM SOC INFORM SCI, V40, P342 | 8 | 6.72 |
| 1990 | | |
| (21) Egghe L., 1990, INTRO INFORMETRICS Q | 24 | 12.06 |

Table 7 shows the most frequently cited publications from authors published in JOI between 1980 and 1990. The proportion for a CR in the table points out the number of citations in percent of the occurrences of all CRs within the same CRY. The user can find this proportion in the column "Percent in Year" in the table of the right side of the program.

Some papers have been cited most frequently because they have been constitutive for a longer-term research program. The paper by White and Griffith (1981) (publication no. 3 in the table), for example, can be considered as the first in a series of studies using "author co-citation analysis" (ACA) at the School of Library and Information Science at Drexel University. Moed, Burger, Frankfort, and van Raan (1985) (publication no. 11 in the table) can be considered as the first and foundational paper for what has grown into the Leiden Centre for Science and Technology Studies CWTS. Callon, Courtial, Turner, and Bauin (1983) (publication no. 7 in the table) has become the *locus classicus* for the introduction of co-word analysis in bibliometrics.

A number of papers are first-authored by Andras Schubert (with Tibor Braun and/or Wolfgang Glänzel as coauthors) of the Information Science and Scientometrics Research Unit (ISSRU) in Budapest which has been a leading center in citation analysis since the early days of scientometrics in the late 1970s. Glänzel and Schubert (1988) (publication no. 18 in the table) is also listed as one of these most frequently cited papers during the 1980s. The paper introduces the so-called "characteristic scores and scales" which are still used and further developed by the first author at the Centre for R&D Monitoring (ECOOM) in Leuven. Cronin's (1984) book (publication no. 10 in the table) and dissertation about the citation process has been a cornerstone to the development of the School of Library and Information Science in Bloomington, Indiana. The list thus maps the institutional groundwork of the field of information metrics during the 1980s.

## 4 Discussion

In this paper, we introduced a new tool – the CRExplorer – which can be used to analyze the CRs of a specific research field or topic. The tool is especially suitable to identify those publications which have been frequently cited by the researchers in a field and thereby to study for example its historical roots. The CRExplorer simplifies the identification of key publications by enabling the user to work with the graph and the list of the CRs; both show the CRs per year and the two representations are related. For example, peaks on the curve in the graph can be clicked and the CRs are then marked which correspond to the peaks in the list. Furthermore, the underlying dataset can be restricted to specific ranges of publication years and a threshold for including CRs can be set.

RPYS can be and has already been used for a variety of research questions. In the Introduction several examples were given how the RPYS has been applied in different contexts. For example, in addition to the historical roots of a research field scientific myths could be identified. Since the CRExplorer allows a flexible work environment using datasets from WoS, the possibilities of analyses are very wide. For example, one could use the tool to inspect the basic literature referenced by a single researcher: Which are the publications a researcher cites most frequently (Leydesdorff, Bornmann, Comins, Marx, & Thor, in preparation)? These publications could be interpreted as his/ her roots or the publications importantly influencing the researcher. The tool could also be used with journal data: Journals in the same fields can be compared in terms of their anchorage in the literature: Do journals have different origins as reflected by the frequently used CRs of the authors (Comins & Leydesdorff, 2016).

Most important is the use of the CRExplorer to reconstruct the evolution of a research topic as a whole and not only to reveal its historical roots (for which RPYS was developed originally). The reference set within the publications of a research topic reflects the totality of "living" papers, i.e. those, which have been cited in the set. The CRExplorer enables to rank these references according to their frequency of occurrence for each reference publication year. By this, the references are normalized in terms of the selected publication set (the references most often refer to papers of the specific research topic analyzed) and normalized with regard to their age (only references each of a specific reference publication year are ranked).

In the era of little science (before around 1950, see Marx & Bornmann, 2010) the number of references of a specific reference publication year within a chosen research topic was comparatively low. Therefore, earlier and most often CRs show up as distinct peaks. In the big science period the strong growth of literature results in numerous references and the highly cited more recent references are usually difficult to identify. However, the table of the CRExplorer can be used to show the top CRs for each referenced publication year.

If a specific research topic is analyzed, the most frequently CRs within the consecutive referenced publication years can be expected to reflect the knowledge development. By this, the time evolution of the scientific discourse can be reconstructed. The papers most important for the community in the course of the time are revealed. However, interpretation is not possible without expert knowledge which has to be combined with the usage of the CRExplorer presented here.

The CRExplorer is not only a helpful tool to analyze a dataset under study, but it also offers a context for cleaning the data. CRs appear in more or less frequent variants; in evaluation studies, for example, it is important to be able to (dis)aggregate the variants before starting with the analysis. Since especially book titles are unstandardized, their role could be undervalued if data were used without first cleaning them. The problem of cited references variants is closely related to the problem of citation linking or matching citations in bibliometrics (Olensky, Schmidt, & van Eck, 2015). Cited references with inaccuracies result in missed matches in the WoS and lead to reduced citation counts for papers (times cited information in WoS). For example, Moed (2005) investigated 22 million cited references from the WoS and found 7.7% discrepant cited references resulting in a missed match with target papers. Thus, solutions are wanted – as offered in the CRExplorer – which contribute to reducing the number of variants of the same cited reference in bibliometric data.

# Acknowledgements

We would like to thank Ludo Waltman, Robin Haunschild, and Jordan Comins for their valuable feedback on an earlier version of our paper and/ or the CRExplorer.

# References

Barth, A., Marx, W., Bornmann, L., & Mutz, R. (2014). On the origins and the historical roots of the Higgs boson research from a bibliometric perspective. *The European Physical Journal Plus, 129*(6), 1-13. doi: 10.1140/epjp/i2014-14111-6.

Bergstrom, C. (2007). Eigenfactor: Measuring the value and prestige of scholarly journals. *College & Research Libraries News, 68*(5), 314-316.

Bornmann, L., Leydesdorff, L., & Mutz, R. (2013). The use of percentiles and percentile rank classes in the analysis of bibliometric data: opportunities and limits. *Journal of Informetrics, 7*(1), 158-165.

Bradford, S. (1934). Sources of information on specific subjects. *Engeneering, 137*, 85-86.

Callon, M., Courtial, J.-P., Turner, W. A., & Bauin, S. (1983). From translations to problematic networks: an introduction to co-word analysis. *Social Science Information, 22*(2), 191-235.

Comins, J. A., & Hussey, T. W. (2015a). Compressing multiple scales of impact detection by Reference Publication Year Spectroscopy. *Journal of Informetrics, 9*(3), 449-454.

Comins, J. A., & Hussey, T. W. (2015b). Detecting seminal research contributions to the development and use of the global positioning system by reference publication year spectroscopy. *Scientometrics*, 1-6. doi: 10.1007/s11192-015-1598-2.

Comins, J. A., & Leydesdorff, L. (2016). Identification of long-term concept-symbols among citations: Can documents be clustered in terms of common intellectual histories? Retrieved January 5, 2016, from http://arxiv.org/abs/1601.00288

Cronin, B. (1984). *The citation process: The role and significance of citations in scientific communication*. Oxford, UK: Taylor Graham.

de Solla Price, D. (1976). A general theory of bibliometric and other cumulative advantage processes. *Journal of the American Society for Information Science, 27*(5-6), 292-306.

de Solla Price, D. J. (1963). *Little science, big science*. New York, NY, USA: Columbia University Press.

Egghe, L. (2006). Theory and practise of the *g*-index. *Scientometrics, 69*(1), 131-152. doi: 10.1007/s11192-006-0144-7.

Einstein, A. (1905a). Über einen die Erzeugung und Verwandlung des Lichtes betreffenden heuristischen Gesichtspunkt [Generation and conversion of light with regard to a heuristic point of view]. *Annalen der Physik, 17*(6), 132-148.

Einstein, A. (1905b). Zur Elektrodynamik bewegter Körper [On the electrodynamics of moving bodies]. *Annalen der Physik, 17*, 891-921.

Garfield, E. (1955). Citation indexes for science - new dimension in documentation through association of ideas. *Science, 122*(3159), 108-111.

Glänzel, W., & Schubert, A. (1988). Characteristic scores and scales in assessing citation impact. *Journal of Information Science, 14*(2), 123-127.

Gonzalez-Pereira, B., Guerrero-Bote, V. P., & Moya-Anegon, F. (2010). A new approach to the metric of journals' scientific prestige: the SJR indicator. *Journal of Informetrics, 4*(3), 379-391. doi: 10.1016/j.joi.2010.03.002.

Guerrero-Bote, V. P., & de Moya-Anegon, F. (2012). A further step forward in measuring journals' scientific prestige: the SJR2 indicator. *Journal of Informetrics, 6*(4), 674-688.

Hazen, A. (1914). Storage to be provided in impounding reservoirs for municipal water supply. *Transactions of American Society of Civil Engineers, 77*, 1539-1640.

Hirsch, J. E. (2005). An index to quantify an individual's scientific research output. *Proceedings of the National Academy of Sciences of the United States of America, 102*(46), 16569-16572. doi: 10.1073/pnas.0507655102.

Lack, D. (1947). *Darwin's finches*. Cambridge, UK: Cambridge University Press.

Leydesdorff, L., Bornmann, L., Comins, J., Marx, W., & Thor, A. (in preparation). Referenced Publication Year Spectrography (RPYS) and Algorithmic Historiography: The Bibliometric Reconstruction of András Schumbert's Œuvre.

Leydesdorff, L., Bornmann, L., Marx, W., & Milojevic, S. (2014). Referenced Publication Years Spectroscopy applied to iMetrics: *Scientometrics*, *Journal of Informetrics*, and a relevant subset of JASIST. *Journal of Informetrics, 8*(1), 162-174. doi: DOI 10.1016/j.joi.2013.11.006.

Lotka, A. J. (1926). The frequency distribution of scientific productivity. *Journal of the Washington Academy of Sciences, 12*, 317 - 323.

Marx, W., & Bornmann, L. (2010). How accurately does Thomas Kuhn's model of paradigm change describe the transition from a static to a dynamic universe in cosmology? A historical reconstruction and citation analysis. *Scientometrics 84*(2), 441-464.

Marx, W., & Bornmann, L. (2014). Tracing the origin of a scientific legend by reference publication year spectroscopy (RPYS): the legend of the Darwin finches. *Scientometrics, 99*(3), 839-844. doi: DOI 10.1007/s11192-013-1200-8.

Marx, W., Bornmann, L., Barth, A., & Leydesdorff, L. (2014). Detecting the historical roots of research fields by reference publication year spectroscopy (RPYS). *Journal of the Association for Information Science and Technology, 65*(4), 751-764. doi: 10.1002/asi.23089.

Moed, H. F. (2005). *Citation analysis in research evaluation*. Dordrecht, The Netherlands: Springer.

Moed, H. F., Burger, W. J. M., Frankfort, J. G., & van Raan, A. F. J. (1985). The use of bibliometric data for the measurement of university research performance. *Research Policy, 14*(3), 131-149.

Olensky, M., Schmidt, M., & van Eck, N. J. (2015). Evaluation of the citation matching algorithms of CWTS and iFQ in comparison to the Web of science. *Journal of the Association for Information Science and Technology*, n/a-n/a. doi: 10.1002/asi.23590.

Page, L., Brin, S., Motwani, R., & Winograd, T. (1998). The pagerank citation ranking: Bringing order to the web. *Stanford Digital Libraries SIDL-WP-1999-0120*.

Pinski, G., & Narin, F. (1976). Citation influence for journal aggregates of scientific publications: theory, with application to literature of physics. *Information Processing & Management, 12*(5), 297-312.

Schreiber, M. (2012). Inconsistencies of recently proposed citation impact indicators and how to avoid them. *Journal of the American Society for Information Science and Technology, 63*(10), 2062-2073. doi: 10.1002/asi.22703.

Schreiber, M., Malesios, C. C., & Psarakis, S. (2012). Exploratory factor analysis for the Hirsch index, 17 h-type variants, and some traditional bibliometric indicators. *Journal of Informetrics, 6*(3), 347-358. doi: 10.1016/j.joi.2012.02.001.

Uzzi, B., Mukherjee, S., Stringer, M., & Jones, B. (2013). Atypical Combinations and Scientific Impact. *Science, 342*(6157), 468-472. doi: 10.1126/science.1240474.

van Eck, N. J., & Waltman, L. (2010). Software survey: VOSviewer, a computer program for bibliometric mapping. *Scientometrics, 84*(2), 523-538. doi: 10.1007/s11192-009-0146-3.

van Eck, N. J., & Waltman, L. (2014). CitNetExplorer: A new software tool for analyzing and visualizing citation networks. *Journal of Informetrics, 8*(4), 802-823. doi: DOI 10.1016/j.joi.2014.07.006.

Vinkler, P. (2010). *The evaluation of research by scientometric indicators*. Oxford, UK: Chandos Publishing.

Wasi, N., & Flaaen, A. (2015). Record linkage using Stata: Preprocessing, linking, and reviewing utilities. *Stata Journal, 15*(3), 672-697.

White, H. D., & Griffith, B. C. (1981). Author Cocitation - a Literature Measure of Intellectual Structure. *Journal of the American Society for Information Science, 32*(3), 163-171. doi: DOI 10.1002/asi.4630320302.

Wouters, P. (1999). *The Citation Culture*. Amsterdam, The Netherlands: Unpublished Ph.D. Thesis, University of Amsterdam.

Wray, K. B., & Bornmann, L. (2014). Philosophy of science viewed through the lense of "Referenced Publication Years Spectroscopy" (RPYS). *Scientometrics*, 1-10. doi: 10.1007/s11192-014-1465-6.

Zhang, L., Rousseau, R., & Glänzel, W. (2015). Diversity of references as an indicator of the interdisciplinarity of journals: Taking similarity between subject fields into account. *Journal of the Association for Information Science and Technology*, n/a-n/a. doi: 10.1002/asi.23487.